\documentclass[preprint2]{aastex}

\def\ltsima{$\; \buildrel < \over \sim \;$}
\def\simlt{\lower.5ex\hbox{\ltsima}}
\def\gtsima{$\; \buildrel > \over \sim \;$}
\def\simgt{\lower.5ex\hbox{\gtsima}}
\def\cgs{{erg cm$^{-2}$ s$^{-1}$}}
\def\ergsc{{erg cm s$^{-1}$}}
\def\ergs{{erg s$^{-1}$}}
\def\cm2{{cm$^{-2}$}}

\def\xrd{{$\chi^{2}_{\rm \nu}$/dof}}
\def\xddof{{$\chi^{2}/$dof}}
\def\xnu{{$\chi^{2}(\nu)$}}

\def\aox{$\alpha_{\rm ox}$}

\def\lums{{$L_{0.5-2 keV}$}}
\def\lum{{$L_{2-10 keV}$}}

\def\xmm{{\em XMM--Newton}}
\def\chandra~{{\em Chandra}}

\def\nhgal{{N$_{\rm H}^{\rm Gal}$}}

\def\chandra{{\em Chandra}}

\def\xmm{{\em XMM--Newton}}
\def\nhgal{{N$_{\rm H}^{\rm Gal}$}}
\def\nh{{N$_{\rm H}$}}
\def\epic{{\em EPIC}}

\def\f14{{10$^{-14}$}}
\def\f13{{10$^{-13}$}}
\def\f12{{10$^{-12}$}}
\def\f11{{10$^{-11}$}}
\def\e22{{10$^{22}$}}

\def\feka{{Fe K$\alpha$}}
\def\3c{{3C 234}}
\def\l58{{$L_{5.8 \mu m}$}}
\def\ftm{{FTM 0830$+$3759}}
\def\red{{\it red}}
\shorttitle{XMM-Newton observations of FTM  0830$+$3759}
\shortauthors{Piconcelli et al.}

\begin{document}

\title{Investigating the complex X-ray spectrum of a broad-line 2MASS red quasar:\\ XMM-Newton observation of FTM~0830$+$3759}

\author{Enrico Piconcelli\altaffilmark{1}, Cristian Vignali\altaffilmark{2},  Stefano Bianchi\altaffilmark{3},
Fabrizio Nicastro\altaffilmark{1,4,5}, Giovanni Miniutti\altaffilmark{6}, and Fabrizio Fiore\altaffilmark{1}}

\altaffiltext{1}{Osservatorio Astronomico di Roma (INAF), Via Frascati 33, I-00040 Monte Porzio Catone, Italy; piconcelli@oa-roma.inaf.it.}
\altaffiltext{2}{Dipartimento di Astronomia, Universit\`a  di Bologna, Via Ranzani 1, I-40127 Bologna, Italy.}
\altaffiltext{3}{Dipartimento di Fisica, Universit\`a  degli Studi Roma Tre, via della Vasca Navale 84, I--00146 Roma, Italy.}
\altaffiltext{4}{IESL, Foundation for Research and Technology, 711 10, Heraklion,  Crete, Greece.}
\altaffiltext{5}{Harvard-Smithsonian Center for Astrophysics, 60 Garden Street,  MS-04, Cambridge, MA 02155, USA.}
\altaffiltext{6}{LAEX, Centro de Astrobiologia (CSIC-INTA) LAEFF, PO Box 78, E-28691 Villanueva de la Ca\~{n}ada, Madrid, Spain.}

\begin{abstract}

We report results from a 50 ks \xmm~observation of the dust-reddened
broad-line quasar \ftm~($z$ = 0.413) selected  from the FIRST/2MASS Red
Quasar survey. For this AGN, a very short 9 ks \chandra~exposure had
suggested a feature-rich X-ray spectrum and {\it HST} images revealed
a very disturbed host galaxy morphology.  Contrary to classical,
optically-selected quasars, the X-ray properties of  {\it red} (i.e.  with $J-K_s>$ 1.7 and $R-K_s>$ 4.0)
broad line quasars are still quite unexplored,
although there is a growing consensus that, due to moderate obscuration, these objects can offer a unique view of spectral 
components typically swamped by the AGN
 light in normal, {\it blue} quasars.
The \xmm~observation discussed here has definitely confirmed the
complexity of the X-ray spectrum revealing the presence of a  cold
 (or mildly-ionized) absorber with \nh~$\approx$ 10$^{22}$ \cm2~along
the line of sight to the nucleus  and  a Compton reflection component
accompanied by an intense \feka~emission line in this quasar with a
\lum~$\approx$ 5 $\times$ 10$^{44}$ \ergs.  A {\it soft-excess}
component is also required by the data.  The match between the column
density derived by our spectral analysis and that expected on the
basis of reddening due to the dust suggests the possibility that both
absorptions occur in the same medium.  \ftm\ is characterized by 
an extinction/absorption-corrected  X-ray-to-optical flux ratio
\aox\ = $-$2.3, that is steeper than expected on the basis of its  UV
luminosity.  These findings indicate that the X-ray properties of
\ftm~differs from those typically observed for optically-selected broad
line quasars with comparable hard X-ray luminosity.
 \end{abstract}


\section{Introduction} 

Selection criteria using near-infrared  and mid-infrared  photometry
or combination of infrared (IR) and multiwavelength data have
demonstrated the striking existence of a conspicuous population of
obscured  active galactic nuclei (AGNs)  that has remained unknown
until fairly recently, being difficult to detect with traditional
selection criteria in the optical band
\citep[e.g.,][]{lacy04,lacy07,leipski05,webster95,fiore08,donley08,lanzuisi09}.
On the one hand, dust extinction hampers a complete census of the
active SMBHs in the Universe in case of optical/UV surveys, on the
other hand,  the   AGN light is absorbed by the obscuring medium and
then re-emitted at near- and mid-IR wavelengths.

Using the Two Micron All Sky Survey (2MASS)
\citep[e.g.,][and references therein]{skrutskie06},  \citet{cutri02} have recently unveiled a
population of highly-reddened ($J - K_s$  $>$ 2) quasars  at an
average redshift of $\langle$$z$$\rangle\sim$ 0.22 whose number density is almost
comparable to that of the  optically-selected quasars,
selected on the basis of the UV-blue color excess criterion
\citep[e.g.,][]{risaliti04, hall06}.  
 Most of these 2MASS sources are previously unidentified quasars
\citep{francis04,  glikman04,leipski05, urrutia09}. Optical follow-up
observations have revealed that the bulk of this extragalactic
population is composed by   broad-line AGNs with a significant number
of intermediate type objects (i.e. Type 1.2--1.9). In particular, it has been
found that: (i) their median spectral energy distribution (SED) is
redder in the optical/UV band than that of the best-studied, classical {\it blue}
quasars \citep[e.g.][]{elvis94}, (ii) they are primarily reddened by
dust (A$_V$ $\sim$ 1--5 mag) rather than being intrinsically red
sources,  (iii) these objects usually exhibit a high optical
polarization level ($\approx$ 3--15\%) suggesting that the
polarization is due to scattering of nuclear light by material located
close to the active nucleus, but exterior to the BLR
\citep{kura09a,smith03}. Such an aspect is very intriguing since
$\sim$20\% of 2MASS AGNs exhibit  the same broad emission lines both in the polarized flux and in the total flux spectrum.
 This indicates that a sizable fraction of the observed
optical AGN emission must have scattered into our direction and, in
turn, suggests caution in the use of spectral type (Type 1 versus Type
2) as indicator of orientation in AGNs \citep{schmidt07}.

The observed properties of \red\ quasars can be
interpreted in terms of
 intermediate viewing angles where the AGN is viewed through 
 the edge (or atmosphere) of the torus or a clumpy accretion disk  wind.
 Accordingly, the observed broad-band emission is the likely combination 
 of direct, reprocessed and scattered emission components \citep[e.g.,][]{smith03,wilkes08}.
 In addition,  as pointed out by \citet{kura09a}, both obscuration and emission from the circumnuclear gas
  and the host galaxy  play a crucial role in shaping the SED of \red\ 2MASS quasars.
 It is also worth noting that a small ($\approx$10--15\%) percentage of classical, {\it blue} quasars is unavoidably 
 collected, especially at $z$ \simlt\ 0.4, by the  $J-K_s>$ 2 selection criterion \citep{barkhouse01}.  Also in \citet{kura09a},
 it is shown that some unreddened (i.e. A$_V$ = 0--1 mag) Type 1 sources with high Eddington ratios ($L/L_{\rm Edd}$)
 and large amounts of hot circumnuclear dust are picked up within their sample of \red\ 2MASS quasars.
 However, since the $J-K_s$ color selection is inhomogeneous with redshift,
 at higher redshifts less contamination from {\it blue} AGNs will occur (as $J-K_s$ color resembles more 
  a rest frame optical color) and the red $J-K_s$ selection will mostly gather AGNs that are truly dust-reddened.

More sophisticated selection processes involving multiwavelength (i.e. radio,
optical, infrared) data have recently
allowed to reveal {\it red} quasars
 (i.e.  with $J-K_s>$1.7 and $R-K_s>$4.0) 
up to $z$ \simgt~1, such as in the case of the VLA FIRST-2MASS (FTM hereafter) 
survey published by \citet{glikman07}.
They estimate that {\it red} quasars account for between 
25\% and 60\% of the total quasar population with $K_s <$ 14 mag.

The nature and the evolutionary properties of {\it red} quasars are still
open issues.
It is worth noting that the classical,
non-\red~AGN population, i.e. with a 2MASS color $J-K_s<$ 2, indeed comprises a mixed bag of objects spanning
 from unobscured {\it blue} quasars
to Seyfert 2-like AGNs showing only narrow lines in their optical spectra and absorbed (even
 with Compton-thick, i.e. \nh\ $>$ 10$^{24}$ \cm2, column densities) at
X-rays \citep[e.g.][]{watanabe04,alonso98,gorjian04,zakamska04}.
In particular, what is the place of {\it red} broad-line quasars in the AGN unification scheme and their role
in the context of AGN and galaxy co-evolution?
Are they standard systems observed along a special line of
sight for which both nuclear and host galaxy obscuration
are important \citep{kura09a}, i.e.
alternative to that of two classical populations of
unobscured broad-lined  quasars and that of heavily obscured,
narrow-lined quasars? Alternatively, can they (especially if they are at $z$ \simgt 0.4--0.5) be interpreted as
a peculiar dust-cocooned stage in the life cycle of quasars, possibly associated with the assembly of the host galaxy? 
 
Intriguingly, HST images available for a sample of dust-reddened FTM
quasars with $M_B$ $\leq$ $-$23 and 0.4\simlt~$z$ \simlt~1 show disturbed
optical morphologies in the vast majority (85\%) of them with evidence
of ongoing merging, interactions and multiple nuclei 
\citep{urrutia08,hutchings06}, unlike {\it blue} quasars being mostly hosted in undisturbed
elliptical or bulge-dominated galaxies as found in HST imaging studies
of  optically-selected luminous quasars at $z$ $<$ 1 
\citep[e.g.,][see also \citealt{bennert08}]{dunlop03,floyd04,
sanchez04}.

Evidence is mounting  that distant {\it red} quasars might
indeed represent  a  dust-enshrouded phase in quasar evolution
linked with the host galaxy assembly via repeated
mergers \citep{georgakakis09,urrutia09}. This phase should be characterized by
the presence of massive nuclear winds expelling/heating most of the
cold gas reservoir and merger debris in the host galaxy, and so
rendering progressively visible the quasar as an optically-bright source
(the so-called {\it feedback} processes, e.g. \citet{silk98,hopkins06}). 


A deep exploration of the X-ray spectral properties of \red~quasars is
therefore important to  improve  our understanding of the structure of
AGNs and their cosmological evolution.  The combination of a
favorable line of sight  inclination and obscuration can indeed provide a unique
view of spectral components usually overwhelmed by the AGN light in
{\it blue} quasars \citep{schmidt07,pounds05}.  The  X-ray spectral
properties of unobscured,  optically-selected, radio-quiet quasars
have been largely investigated in the past
\citep{laor97,george00,pico05}, especially in case of
medium-luminosity objects showing an X-ray luminosity of L$_X$ $<$ 5
$\times$  10$^{44}$ \ergs. They appear to be quite  homogeneous and
without any evidence for significant  evolution with $z$
\citep{just07,page05, pico03,vignali03a}, although  the strength of
spectral features due to the reprocessing of the primary continuum
(such as \feka~emission line and Compton hump) is still largely
unconstrained for high luminosity quasars with L$_X$\simgt~10$^{45}$
\ergs~\citep{mineo00,page05,jimenez05}.

On the contrary, the X-ray properties of  {\it red} luminous quasars
still remain poorly explored. After their discovery, a handful of
programs aimed at studying the X-ray properties of these AGNs have
been undertaken.  \chandra~\citep[hereafter
U05]{wilkes02,hall06,urrutia05} and \xmm~shallow observations
\citep{wilkes05,pounds05} of $\sim$20-30 2MASS quasars indicate  the
quasi-ubiquitous presence  of cold absorption with
10$^{21}$\simlt\nh\simlt 10$^{23}$ \cm2~regardless of optical type.
This matches with the finding that only $\approx$10\% of the \red~2MASS
AGNs are detected in the ROSAT Faint Source Survey \citep{cutri02}.

Bearing in mind the small sample size to date, broad-line \red~quasars seem to
show a wide range of spectral types in the X-ray band unlike their
{\it blue} counterparts. In particular, two results from
X-ray spectroscopy of \red~quasars can be considered of special
interest: (i) a number of these sources have an unusually  flat hard X-ray
continuum with $\Gamma_{2-10}$$<$1.5
\citep[e.g.][]{vignali00,wilkes05,pounds07,wilkes08},  probably due to
the presence of an intense Compton reflection component, and (ii)
some broad-line \red~AGNs exhibit a "soft-excess" component below
$\sim$1-2 keV that can be interpreted as the result of absorption and
re-emission from the same weakly ionized (outflowing) gas
\citep{pounds05,pounds07}. 

In this paper, we present  the first high-quality X--ray spectrum,
obtained with \xmm,  of the luminous  red quasar
\ftm~at $z$=0.413.  A very short 9 ks {\em Chandra} observation
revealed a very steep continuum and the possible presence of puzzling
emission/absorption line-like features at energies which do not
correspond to any  obvious rest-frame atomic transition (U05).  We assume
throughout this paper $H_0$ = 70 km s$^{-1}$ Mpc$^{-1}$,
$\Omega_\Lambda$ = 0.73 and  $\Omega_M$= 0.27 \citep{spergel07}.

%
%
\begin{deluxetable}{ccccccc}
\tabletypesize{\footnotesize}
\tablecaption{\label{f2m}Multiwavelength properties of \ftm} 
\tablewidth{0pt}
\tablehead{
 \colhead{R.A.} &  \colhead{Dec.} &  \colhead{$z^a$} & \colhead{Flux 1.4 GHz$^a$} &  \colhead{$J - K^{b}$}& \colhead{$E(B-V)^{c}$} & \colhead{M$^{c}_B$}  \\
 \colhead{($h\,m\,s$)} &  \colhead{($\circ\,'\,''$)} & \colhead{} & \colhead{(mJy)}  &\colhead{(mag)}  &\colhead{(mag)} &  \colhead{(mag)}
\label{Obstab}}
\startdata
 08\,30\,11.12 & $+$37\,59\,51.8 & 0.413 & 6.4 & 2.28 &1.29$\pm$0.12 & $-$23.07 \\
\enddata
\tablecomments{All magnitudes are in the Vega system.}
\tablenotetext{a}{Data from \citet{urrutia05}.}
\tablenotetext{b}{Data from \citet{glikman07}.}
\tablenotetext{c}{Nuclear values from \citet{urrutia08} based on HST/ACS Wide Field Camera observations.}
\end{deluxetable}

\section{Optical and Radio Properties of  FTM 0830$+$3759}

\ftm~\citep[$J-K$ = 2.28, e.g.][]{glikman07} is the X-ray brightest object
included in the U05 sample of 12
broad-line,  dust-reddened quasars selected from the FTM
survey. The objects in this sample need to satisfy three main
criteria: being a FIRST radio source; belonging to the 2MASS
point-source catalog and having a $R-K$ \simgt~4.4 (we refer to U05
for further details on the optical/near-IR selection process).  All
these luminous \red~quasars lie in the redshift range 0.4\simlt $z$\simlt 2.7, exhibit  broad lines in
their optical spectra 
and they are radio-quiet or radio-intermediate,
i.e. with a ratio of radio to optical emission R$_L$=$F_{\rm
5GHz}/F_{\rm B}<$100. The value of  R$_L$ derived for \ftm~is
$\approx$ 30, and the steep radio spectral index ($\alpha_{1.4 GHz/8.3 GHz}$ = $-$1.06)
derived by \citet{glikman07} also supports the moderate radio-loud nature of this quasar.

\ftm~is also included in the HST study of a sample of 13 dust-reddened
FTM quasars performed by \citet{urrutia08}.  HST ACS Wide Field Camera
images reveal that the host galaxy of \ftm~shows a lot of
irregularities near the nucleus along the major axis.  A prominent
structure with a ionization cone morphology is present: it can be
interpreted as due to merger-induced star formation regions or to the
presence of outflowing gas.  After an accurate PSF
fitting  and host galaxy subtraction,  \citet{urrutia08} measured a
M$_B$ = $-$23.07 mag and a reddening of $E(B-V)$ = 1.29 for the quasar
in \ftm. The inferred value of the nucleus to host ratio for the $I$
magnitude (F814W filter) was 0.55.  Interestingly,   \citet{urrutia08}
also found a possible correlation between the magnitude of galaxy interactions
and the level of obscuration affecting the
quasar. This led them to provide a possible explanation of the red
nature of FTM quasars in terms of reddening occurring in the host
galaxy.

Finally, there is no detailed information on the optical classification of \ftm\ in the
 literature besides it is a broad-line AGN. 
 However, the optical spectrum reported in Fig. 1 of \citet{urrutia08} suggests that an 
 intermediate (i.e. Type 1.5) classification for 
this source may be considered plausible, although a proper investigation on this issue is required.

\section{XMM-Newton Observation and Data Reduction}

We observed the quasar \ftm~with \xmm~\citep{jansen01} on November 8,
2008 for about 52 ks (Obs. ID.: 0554540201).  The observation was
performed with the {\it EPIC } PN and MOS cameras operating in
Full-Window mode and with the THIN filter applied.  Data were reduced
with SAS v8.0.0 using standard procedures
and the most updated calibration files available at the date of the
analysis (April 2009) were used. X-ray events corresponding to
patterns 0--4(0--12)  for the PN(MOS)~cameras were selected. The event
lists were filtered to ignore periods of high background flaring
according to the method presented in \citet{pico04} based on
the cumulative distribution function of background lightcurve
count-rates.  After screening the final net exposure times were 41 and
48 ks for PN and MOS, respectively.

The source photons were extracted for the PN(MOS) camera from a
circular region with a radius of 29(30) arcsec centered at the peak of
the emission, while the background counts were estimated from a
52(49) arcsec radius source-free region on the same chip and close to
\ftm~without being contaminated by the target itself.
 
The redistribution matrix files and ancillary response files were
created using the SAS task RMFGEN and ARFGEN, respectively.  As the
difference between the MOS1 and the MOS2 response matrices is a few
percent, we created a combined MOS spectrum and response matrix. The
background-subtracted spectra for the PN and the combined MOS cameras
were then simultaneously fitted. Spectra were rebinned so that
each energy bin contains at least 20 counts to allow us to use the
$\chi^2$ minimization technique in spectral fitting.

During the \xmm~observation the flux of \ftm~remained steady, with
no variation exceeding 2$\sigma$ from the average count-rate level
in both soft- and hard-X-ray band. Since no significant spectral
changes occurred, the spectral analysis was performed on the
spectrum integrated over the full exposure time.

\section{Spectral Analysis} 
 In this Section we present the spectral
analysis of the \epic~observation of \ftm~ that was carried out using
the XSPEC v11.3 software package \citep{arnaud96}.  The Galactic column
density of \nhgal\ = 3.61 $\times$ 10$^{20}$ \cm2~derived from
\citet{kalberla05} was adopted in all the fits. In the following,
errors correspond to the 90\% confidence level for one interesting
parameter, i.e. $\Delta\chi^2$= 2.71 \citep{avni76}.

\subsection{Preliminary Fits}

\begin{figure*}
\begin{center}
\includegraphics[width=6.5cm,height=8.5cm,angle=-90]{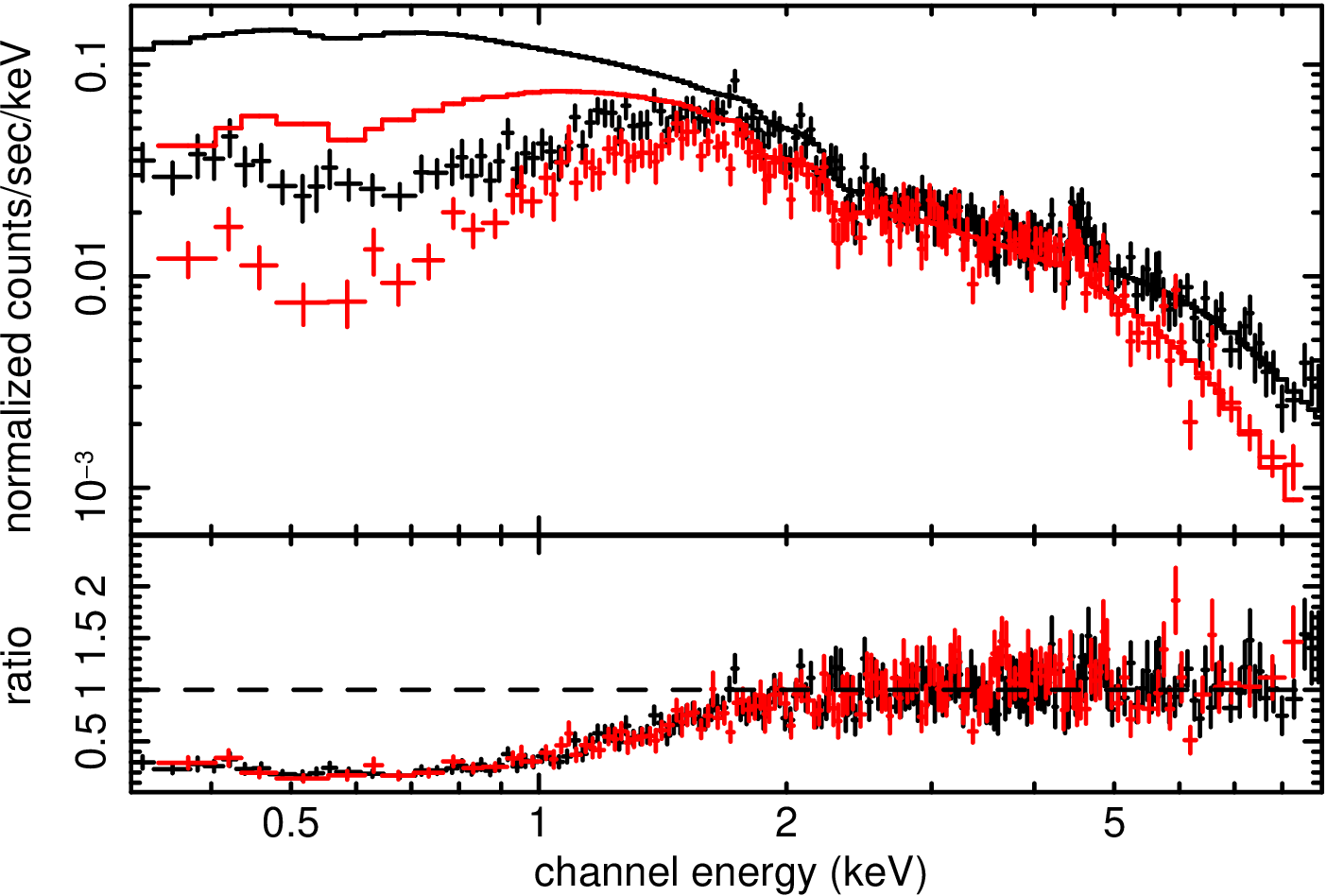}\hspace{0.2cm}\includegraphics[width=6.5cm,height=8.5cm,angle=-90]{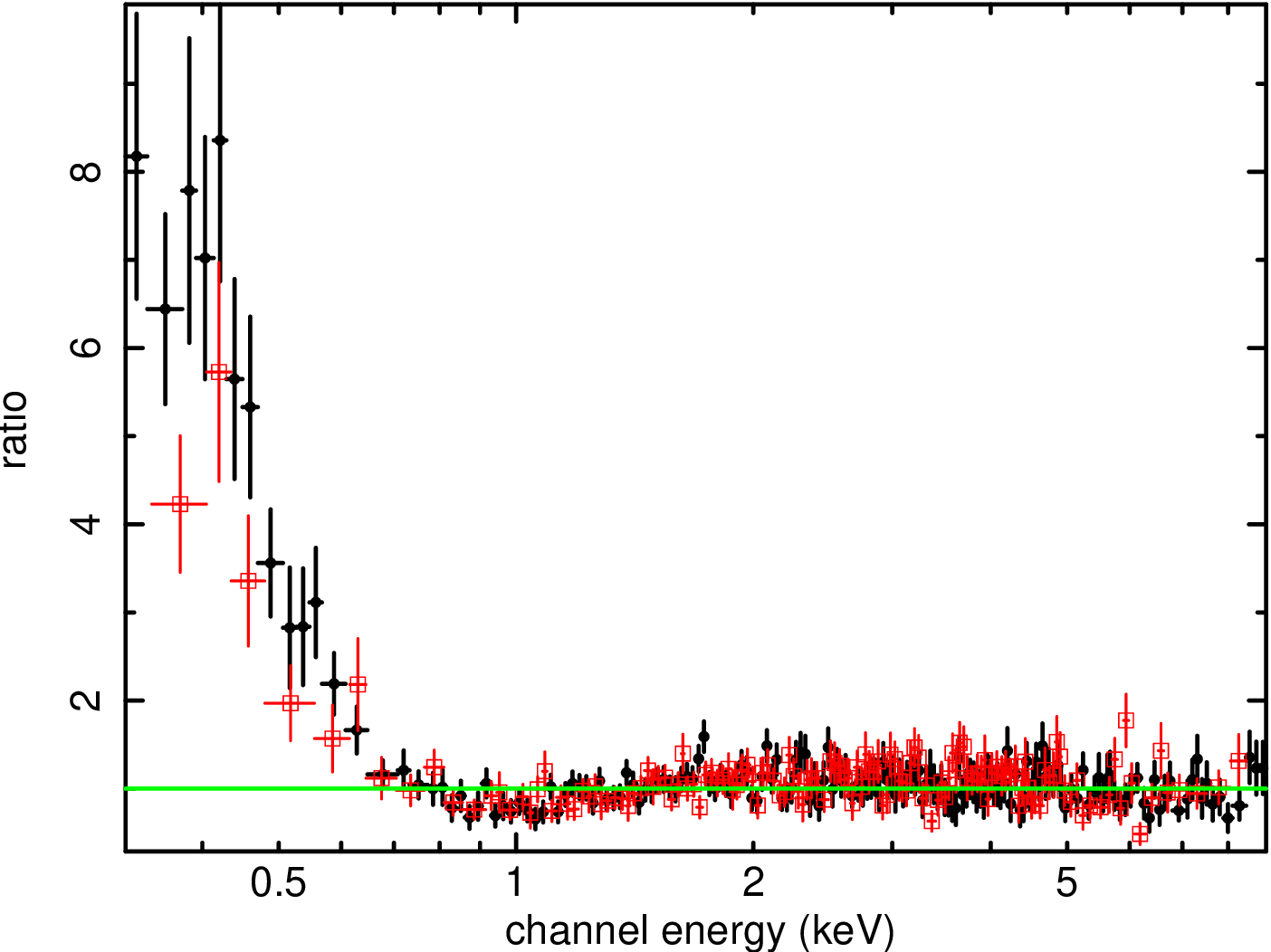}
\caption{{\it Left:} 
Continuum power-law fit ($+$ a Gaussian emission line at $\sim$ 6.4 keV rest-frame) to the 2.5-9 keV band of the PN ({\it top})  and MOS  ({\it bottom}) spectra of \ftm~extrapolated over the 0.3-9 keV band.
 The lower panel shows the data/model ratio residuals.
 {\it Right:} The data/model ratio residuals resulting from fitting an absorbed power law to the 0.3-9 keV EPIC data 
 (circles and empty squares are the PN  and MOS residuals, respectively).
The shape of the residuals strongly suggests the presence of a {\it soft-excess} emission component below $\approx$0.7 keV.}
\label{fig:12}
\end{center}
\end{figure*}

As a first step we fitted the hard
portion of the \epic~spectra at $E$$>$2.5 keV (corresponding to 3.5
keV in the source frame) with a  power law  to achieve a
preliminary description of the X-ray primary continuum emission in an energy range expected to be much less affected by
absorption than soft X-rays. The  emission line, observed at $\sim$ 4.5 keV (6.4 keV in the quasar frame), was modelled
with a narrow Gaussian line with energy and normalization free to
vary (see Sect. 4.3 for more details). This model is a good fit to the \epic~data,
i.e. \xrd~= 1.02/174.  We measured a $\Gamma$ = 1.37$\pm$0.08 that is
significantly flatter than the average hard X-ray slope
$\langle\Gamma\rangle$ $\approx$ 1.85 found for optically-selected
quasars \citep{pico05}.  Fig.~\ref{fig:12} (left panel) shows the result of this
simple fit extrapolated to 0.3 keV. The broad and deep deficit in the
soft X-ray band clearly suggests the presence of strong
absorption. Fitting the data over the 0.3-9 keV band with an
unrealistic power-law model yielded  a $\Gamma \sim$0.72.

To reproduce the shape of the soft X-ray spectrum we initially applied a
model consisting of a power law modified by intrinsic, rest-frame
cold matter absorption. This fit was statistically unacceptable with
\xrd~= 1.95/324, revealing additional spectral complexity.
In particular, an excess in the data/model residuals emerges at energies
 below 0.6 keV (e.g. Fig.~\ref{fig:12}, right panel). 

To further explore the nature of the obscuring material along our line
of sight we modeled the low-energy drop by replacing the 
neutral absorber with a photoionized absorber component.
The latter was modeled in XSPEC adopting the publicly available output table 
``grid 18''\footnote{See http://heasarc.gsfc.nasa.gov/docs/software/xstar/xstar.html for further details} 
of the XSTAR code  \citep{kallman01}.
In this model the state of the warm
absorber is a function of the ionization parameter $\xi$ defined as
$\xi$ = $L$/$n$$r^{2}$, where $L$ is the isotropic luminosity of the power-law
ionizing source in the interval 13.6 eV to 13.6 keV, $n$ is the 
density of the plasma  and $r$ represents the radial distance from the central source. In the fit solar elemental abundances
were assumed and  $\xi$ was left as a free parameter. 
 However, this fit, with \xrd=1.05/323, was not entirely effective in
reproducing the \epic~data as broad positive residuals were still present between 0.5 and 0.7 keV.

\subsection{Complex models}
\label{complex}

In order to account for the {\it soft-excess} component,
we then included an additional unabsorbed power-law
 fixing its photon index to that of the absorbed power law but with a different normalization.
Such a parametrization typically provides an excellent description of
the X-ray spectrum of many well-studied Compton-thin AGNs, in which the {\it soft-excess} is
due to a combination of emission from scattered continuum photons and distant photoionized gas
\citep[e.g.][]{turner97, winter09, derosa08}.
 This fit resulted in a \xrd\ =~0.99/323
yielding a dramatic improvement in fit quality over the simple
absorbed power-law model at the $>$99.99\% confidence in an $F-$test.
 We measured
a photon index of $\Gamma$ = 1.51$\pm$0.06 and a column density   of \nh~= (1.86$\pm$0.16) $\times$
10$^{22}$ \cm2~of the neutral absorber, while the ratio between the normalization of the
unabsorbed and  absorbed PL is $f_s$ = 0.13$\pm$0.02.
Table 2 lists the best-fit values of the relevant spectral parameters
obtained by this fit, hereafter referred as NP model.

A  warm absorber model (WP) was then tested, using again the ``grid 18'' XSTAR table.
This fit yielded
\xrd~= 0.97/322 with an improvement at $>$95.6\% confidence level
according to an $F-$test once compared with the NP model.  We derived
the following best-fit values for  the physical parameters of the warm
gas:   a column density of \nh~$\sim$ 2.4 $\times$ 10$^{22}$ \cm2~and
an ionization parameter of $\xi$ $\sim$ 15 \ergsc.

We note that leaving the photon index of the unabsorbed power law free
to vary  in both NP and WP models gave values that are consistent with
the photon index of the continuum and did not produce any appreciable
improvement in the quality of the fits.\\

 It is worth noting that a fitting model with an
absorber completely covering the nucleus plus an extra-continuum X-ray
emission component (like both NP and WP model) is numerically
equivalent to a model with a  partially-covering absorber.
 In the partial covering scenario the soft X-ray emission is alternatively interpreted as a portion
of primary radiation leaking through the absorber with a covering fraction equal to $C_f$ = (1 $-$ $f_s$); see Table 2.
Instead, in case of
fully-covering absorber, the {\it soft excess} is explained in terms of
emission due to reprocessing of the nuclear continuum by surrounding
material.  This reprocessed component may be electron-scattered
emission by highly ionized matter extending above the ``hole'' of the
{\it torus} or Compton reflected by cold material, and/or it may arise
from large-scale ($\sim$0.1-1 kpc; see \citet{bianchi06,bianchi07a})
photoionized gas as pointed out in most of high-resolution soft X-ray
spectra of heavily-obscured Type 2 AGNs which are  dominated by a
wealth of strong emission lines from hydrogen- and helium-like ions of
the most abundant metals, from carbon to sulfur
\citep{guainazzi07,kinka02}.

 Unfortunately, the combination  of \epic~resolution, intermediate
column density of the X-ray absorber and moderate soft X-ray flux of
\ftm~heavily hampers the detection of these emission lines in  the
\xmm~spectrum presented here.  This implies that we were not able to
discriminate between the two proposed scenarios for the {\it soft-excess}
emission and, in turn, provide unambiguous constraints on the covering
fraction of the absorber in \ftm. This aspect should be borne in mind
while interpreting all the results from the spectral analysis
presented hereafter.
Furthermore, we cannot use the high resolution RGS data to shed light on this issue since
\ftm~is too faint at soft X-rays to be detected by the RGS with
sufficient signal to allow a useful spectral analysis.\\

\begin{deluxetable}{cccc}
\tabletypesize{\footnotesize}
\tablecaption{\label{param}Best Fit Continuum Parameters}
\tablewidth{0pt}
\tablehead{
  \colhead{Model} & \colhead{Model Name} & \colhead{Model Parameters} & \colhead{$\chi^{2}$/d.o.f.}
          }
\startdata
APL (neutral absorber) $+$            & NP & \nh = ($1.86 \pm 0.16$) $\times 10^{22}\; {\rm cm}^{-2}$; $f_s$ = 0.13$\pm$ 0.02           & 321/323\\
PL             &  & $\Gamma$ = $1.51 \pm 0.06$                         & \\
\\

 APL (warm absorber) $+$    & WP & \nh~= ($2.40^{\;+0.15}_{\;-0.18}$) $\times 10^{22} \; {\rm cm}^{-2}$; $f_s$ = 0.10$\pm$ 0.04  &  313/322\\
PL      &  & log($\xi$) = 1.20$\pm$0.11 \ergsc; $\Gamma$ = 1.48$\pm$0.05         &\\
                      \\

NP + Compton    & refl-NP & \nh~= ($1.94 \pm 0.15$) $\times 10^{22} \; {\rm cm}^{-2} $; $f_s$ = 0.11$\pm$0.01       & 312/323\\
Reflection (neutral)  &  & $\Gamma$ = 1.67$\pm$0.06             &\\
\citep{magdziarz95}                       &  &   $R$ = $\Omega/2\pi$ $\equiv$ 1      &\\
\\
WP + Compton    & refl-WP & \nh~= ($2.45 \pm 0.15$)$ \times 10^{22}\; {\rm cm}^{-2}$; $f_s$ = 0.09$\pm$0.03     & 307/322\\ 
Reflection (neutral)  &  & log($\xi$) = $1.16^{\;+0.11}_{\;-0.16}$ \ergsc; $\Gamma$ = $1.64^{\;+0.06}_{\;-0.08}$              &\\
\citep{magdziarz95}        &  & $R$ = $\Omega/2\pi$ $\equiv$ 1                   &\\
\\
Ionized absorber/emitter + PL +&refl-wabs  & \nh~= ($1.26 \pm 0.11$)$ \times 10^{22}\;{\rm cm}^{-2}$;  $f_s$ = 0.07 $\pm$ 0.02  & 309/321\\
 Compton Reflection (neutral)                      &  &log($\xi$) = $0.79^{\;+0.18}_{\;-0.11}$ \ergsc; $\Gamma$ = $1.52 \pm 0.03$           &\\
 \citep{magdziarz95}                       &  &     $A_{\rm Fe}$ $\equiv$ 3; $R$ = $\Omega/2\pi$ $\equiv$ 1                           &\\
\\
refl-NP + thermal& th-NP & \nh~= ($8.50 \pm 0.17$)$ \times 10^{22}\;{\rm cm}^{-2}$; $f_s$ = 0.11 $\pm$0.01 & 303/321\\
emission \citep{kaastra96}                      &  & $\Gamma$ = $1.64 \pm 0.06$;  $R$ = $\Omega/2\pi$ $\equiv$ 1              &\\
                      &  & kT\tablenotemark{a}  $<$0.13 keV                             &\\
\enddata
\tablecomments{APL: absorbed power law. PL: unabsorbed power law.}
\tablenotetext{a}{The thermal emission {\tt mekal} component has a 0.5--2~keV
unabsorbed luminosity of ($2.1$) $\times 10^{42}$ \ergs~and contributes $<$1\% of the flux in the 0.5--2~keV band.}
\end{deluxetable}

\begin{figure*}
\begin{center}
\includegraphics[width=6.5cm,height=8.5cm,angle=-90]{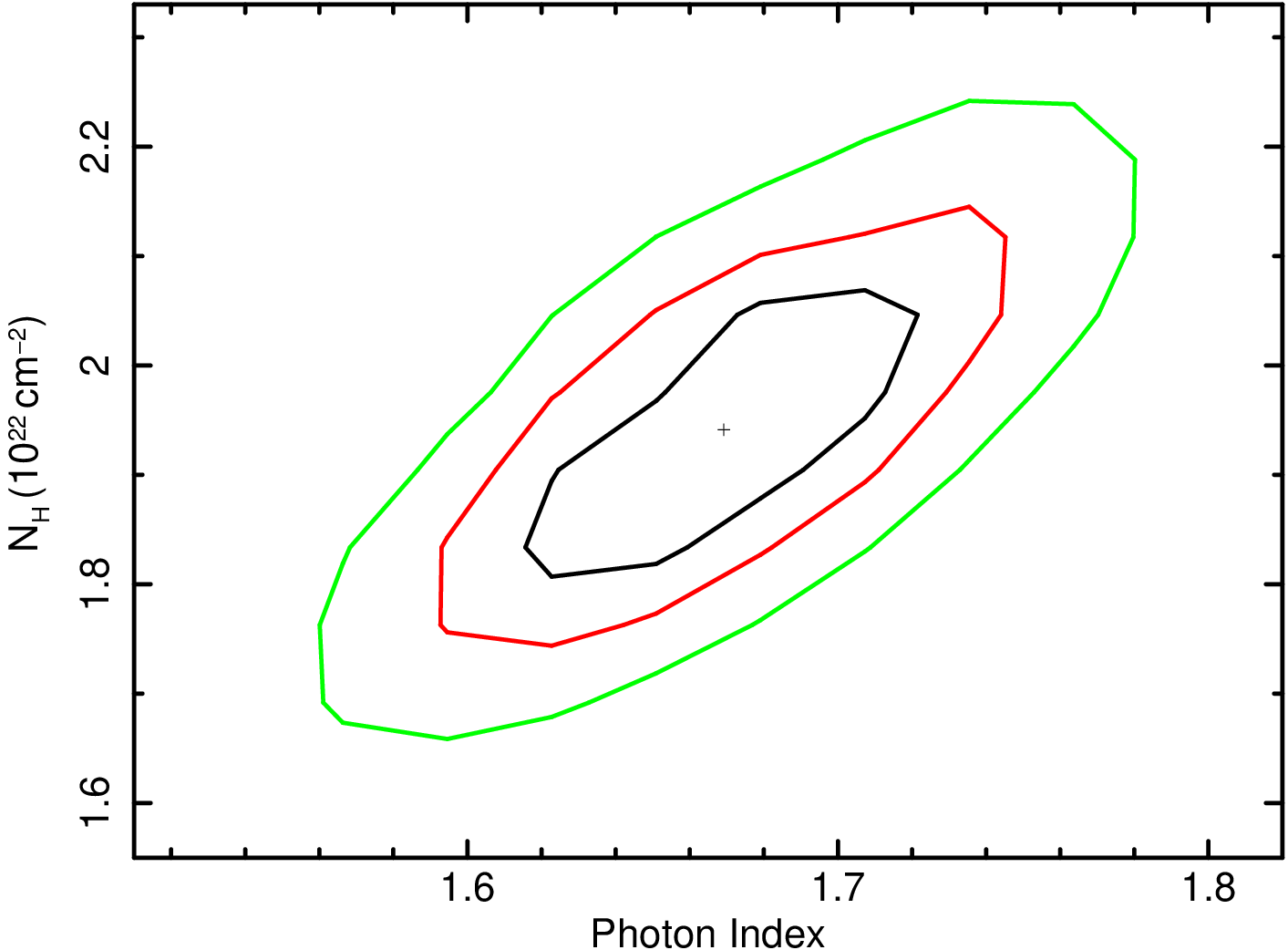}
\hspace{0.2cm}
\includegraphics[width=6.5cm,height=8.5cm,angle=-90]{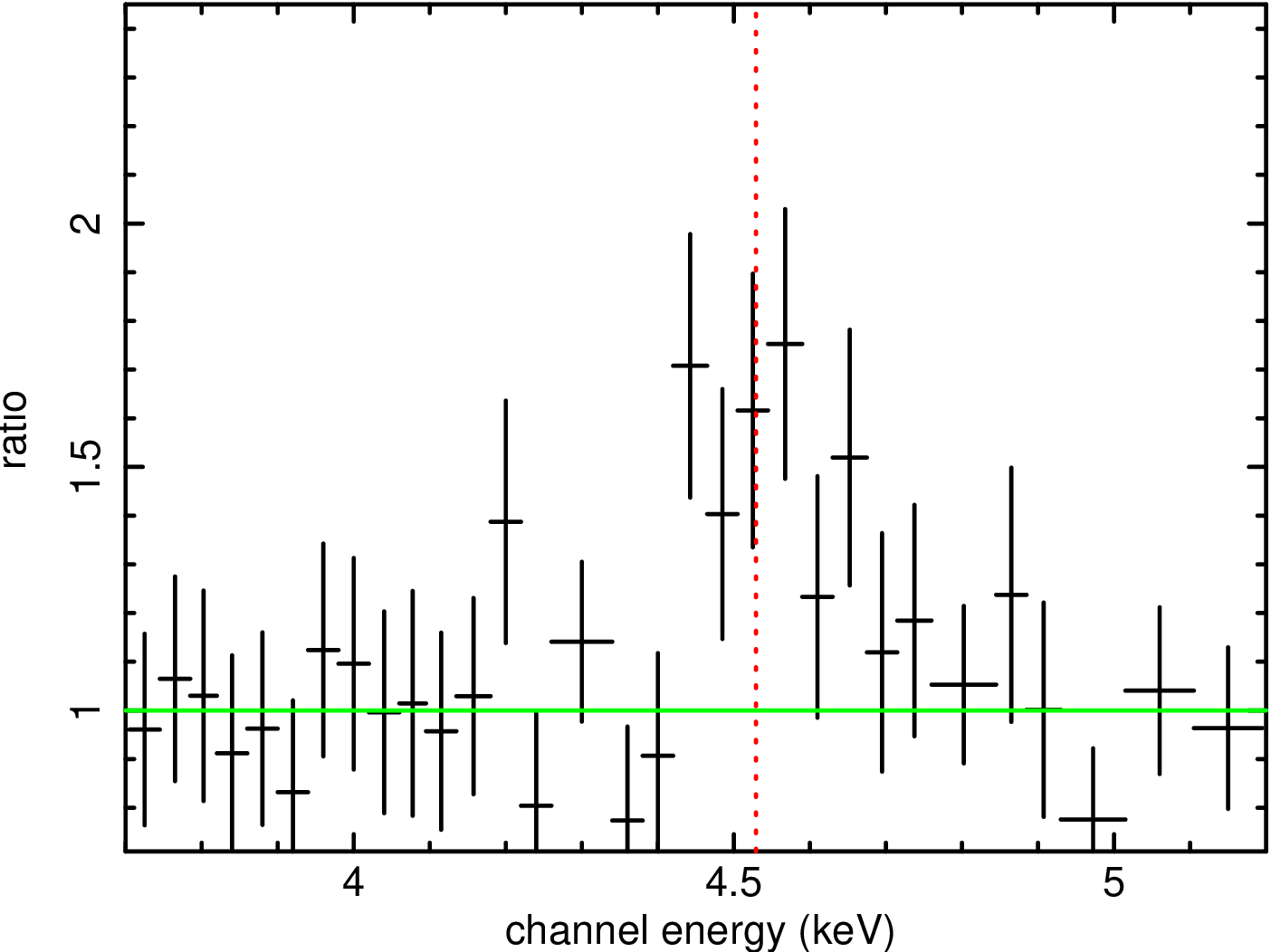}
\caption{{\it Left:} Confidence contour plot showing the column density (in units of 10$^{22}$ \cm2) of the cold absorber against the photon index of the power-law continuum obtained by applying model refl-NP (e.g. Table 2, Sect. \ref{complex}). The contours are at 68\%, 90\%, and 99\% confidence levels for two interesting parameters, respectively.
{\it Right:} Close-up of the \epic~PN data/model ratio residuals in the Fe K$\alpha$ emission line energy range (observer-frame)
when fitting the refl-NP model over  0.3-9 keV and ignoring the
3.8-5.2 keV band. The vertical dashed line represents the energy of 6.4 keV
in the quasar rest-frame.}
\label{fig:34}
\end{center}
\end{figure*}

Given the flatness of continuum slope measured with the NP and WP
models, i.e. $\Gamma \sim$1.5,  we attempted to explain it as the
result of a combination of an  underlying steep continuum modified by a
cold/warm  absorber plus a Compton reflection component from neutral
matter ({\tt pexrav} model in XSPEC, \citet{magdziarz95}), with the
metal abundances of the reflector fixed to the solar value and its
inclination angle fixed to 65 deg. The addition of this spectral
component produced a steeping of the photon index to $\Gamma$
$\approx$1.85-1.9.  However the measured
value of the reflection fraction $R$ $\sim$ 3 (defined as
$\Omega$/2$\pi$, where $\Omega$ is the solid angle subtended by the
reflector for isotropic incident emission) is largely unconstrained
and implies an unphysical covering factor of neutral reflecting
material of $C_f$$>$ 1 and/or anisotropy in the irradiation.  Such a
value is well outside the range typically observed for radio-quiet
broad-line AGNs, i.e. 0.6\simlt~$R$ \simlt~1.2 \citep{perola02,
deluit03, nandra07, panessa08}.  A fit with $R$ fixed at unity resulted
in a \xnu~similar to that obtained with $R$ left free to vary, 
then we decided to use $R$$\equiv$1 in the following fits.  The addition of
a reflection component in the NP(WP) model caused $\chi^2$ to decrease
by 9(6), see model refl-NP(refl-WP) in Table 2. The photon index
steepens to $\Gamma$ $\approx$1.65 in both absorption scenarios
(e.g. Fig.~\ref{fig:34} (left panel) for the iso-$\chi^2$ contour
plot of \nh\ versus $\Gamma$ for the refl-NP model).  Given the large
luminosity of \ftm~we also tested the possible presence of a
reflection component from ionized material by replacing the {\tt
pexrav} component in the refl-NP and refl-WP model with the {\tt
reflion} model of \citet{ross05}. This model is self-consistent
as it  also incorporates fluorescent emission lines from ionized
species. Accordingly we removed from
the fitting model the Gaussian line used in all the previous fits to
describe the line-like feature around 6.4 keV (quasar-frame).  The best-fit values of the
ionization parameter   of the reflector were found to
be consistent with the minimum permitted value of $\xi_{refl}$ = 30
\ergsc~(with an upper limit of $\xi_{refl}$ \simlt\ 40 \ergsc), which
yielded  \xrd~values slightly worse than those with the {\tt pexrav}
component. This clearly indicates that the reflection medium is
almost neutral. Using the formula (1) reported in \citet{brenneman09},
we were able to provide an estimate of the reflection fraction
$R$. Regardless the physical state of the absorber,  we measured an
$R$ value of $\approx$0.6,  that supports a scenario in which the hard
X-ray spectrum of \ftm~is not largely reflection-dominated. 
 

Furthermore,  we examined the intriguing possibility that  the {\it
soft-excess} could instead be the result of  a combination of scattered
continuum and multiple emission lines arising from the same mildly
ionized gas responsible of the absorption
\citep[e.g.,][]{pounds05,mckernan07,longinotti08}. To model this soft
X-ray emitter we again used the XSTAR output table grid 18, with the
ionization parameter $\xi$ and the column density of the emitter tied
to the values of the absorber. We also added to this model a neutral
reflection component with $R$ fixed to 1.  This model (e.g. refl-wabs
in Table 2) gave a very good fit to the \epic~data with \xddof~= 309/321
when the iron abundance was fixed to $A_{\rm Fe}$$\equiv$3 (i.e. its best-fit value) for the
warm gas.  
We deduced a low ionization
parameter of $\xi$ $\approx$ 6 \ergsc~and a column density of
\nh~$\approx$ 10$^{22}$ \cm2~for the absorbing/emitting
gas. The photon index of the power law resulted $\Gamma$ =
1.52$\pm$0.03, i.e. flatter than that measured with the refl-WP model
($\Gamma\approx$1.65).

Finally, we considered a model where a thermal plasma component
({\tt mekal} in XSPEC) is responsible for a fraction of the soft X-ray
positive excess shown in Fig. \ref{fig:12} (left panel), under the
hypothesis of an origin from  starburst activity.  We therefore
included a {\tt mekal} component \citep{kaastra96}   with solar
abundance  in the refl-NP model.  The resulting \xddof~was excellent
with 303/321, significant at 99.1\% confidence in an $F-$test
(e.g. model th-NP in Table~\ref{param}).  However, the best-fit
temperature of the emitting plasma was pegged at the minimum value
allowed of k$T$ $\approx$ 0.08 keV with an upper limit of 0.13
keV.  It is also important to note that the derived 0.5-2 keV luminosity of 2.1 $\times$ 10$^{42}$ 
\ergs~for this thermal component even
exceeds the highest values  measured for starburst regions in
ultra-luminous infrared galaxies \citep[e.g.,][]{franceschini03}.

\subsection{The  Iron K$\alpha$ emission line}

The presence of an excess around 6.4 keV (rest-frame) in the
\xmm~spectrum of \ftm~is clearly evident.  Fig.~\ref{fig:34} (right panel) shows
the  data/model ratio in the observer Fe K emission line energy range
when fitting the refl-NP model over  0.3-9 keV and ignoring the
3.8-5.2 keV band.

We have modeled this spectral feature with a narrow Gaussian line. The
inclusion of this component in the fit  resulted in very significant
statistical improvement ($\Delta\chi^2$ = 13 for two additional free
parameters), i.e.  at 99.99\% confidence level according to an
$F$--test once compared to a model without the line. We concentrated
only on PN data for a more accurate determination of the \feka~line
parameters since the PN CCD has a better sensitivity and energy
resolution over the 4-6 keV energy range  than the MOS CCD.  The
rest-frame energy centroid of the  line is $E$ = 6.42$^{+0.04}_{-0.06}$
keV. This energy indicates a low ionization state, Fe I--XVIII, of the
emitting material. Allowing the width of the line in \ftm~to vary did
not improve the fit over a narrow-line model ($\sigma$\simlt~0.18
keV).  We measured a redshift-corrected equivalent width EW = 168$\pm$60 eV 
assuming
that the line and continuum are both absorbed,
 while the EW calculated with respect to the reflected
continuum is EW = 800$\pm$280 eV.

\subsection{X-ray Fluxes and Luminosities of  \ftm}
\label{fluxes} Once model refl-NP was assumed, we measured an
observed(intrinsic) 2-10 keV flux of 8.9(9.4) $\times$ 10$^{-13}$
\cgs\ and a 0.5-2 keV flux of 1.3(4.1) $\times$ 10$^{-13}$ \cgs.
After correction for both Galactic and  intrinsic absorption, the
luminosity of \ftm\ are \lum\ = 4.8 $\times$ 10$^{44}$   and \lums\ =
2.3 $\times$ 10$^{44}$ \ergs\ in the hard and soft band,
respectively.  The average bolometric corrections measured for
radio-quiet AGNs with similar \lum\ from \citet{vasudevan07}  are in
the range 20--40, that imply a bolometric luminosity of $\approx$
0.97--1.94  $\times$ 10$^{46}$  \ergs\ for \ftm.

\section{Chandra Observation}
\subsection{Data Reduction} 
U05 presented the results of a  \chandra~ACIS-S snapshot survey of
 broad line \red\ quasars aimed at measuring the gas-to-dust ratios of the medium
obscuring the nucleus. 
\ftm, the brightest object in this
 sample, was observed by \chandra~for
$\approx$9 ks in January 2004.

Event files of the {\it Chandra} observation were retrieved from the {\it Chandra}
X-ray Center. The data reduction was performed with the CIAO
v4.1 package following the standard procedures outlined in the Science
Analysis Threads for ACIS data at the CIAO Web site.  Source and
background spectra were produced using the {\it
psextract\footnote{http://cxc.harvard.edu/ciao/threads/psextract/}}
script.  For the creation of the response matrices with the newest
calibrations available we used  the {\it mkacisrmf} task.  The ACIS
source spectrum in the 0.3-7 keV band was grouped to a minimum of 20
counts per bin (see Fig.~\ref{fig:56}).

\subsection{Spectral Analysis}

U05 fitted the spectrum of this quasar with an absorbed, steep power-law ($\Gamma$
$\sim$ 2.9; \nh~$\sim$ 2.7 $\times$ 10$^{22}$ \cm2) plus a broad
($\sigma\sim$ 0.6 keV) Gaussian emission line at 6.4 keV. However, they
also reported the presence of evident unfitted absorption/emission
features.  We have therefore re-analyzed  those data in order to
provide a better characterization of the X-ray spectrum of this
powerful quasar.
Our analysis  of the \chandra~data of \ftm~resulted quite different
from that published by U05.  A fitting model similar to the NP model described in Sect. \ref{complex} with
\nh~$\approx$ 2.1 $\times$ 10$^{22}$ \cm2\ and $f_s$ $\approx$ 0.07 
provided a very good description with a final \xddof=38/36. We
measured a power-law photon index of $\Gamma$ = 1.65$\pm$0.25,
i.e. significantly flatter than that reported by U05 and 
consistent with that obtained from the \xmm~observation.  An
alternative fit with a warm absorber was slightly worse,  yielding
\xddof~= 42/36 (\nh~$\approx$ 2.5$\times$ 10$^{22}$ \cm2~and
ionization parameter $\xi$ $\approx$  23 \ergsc). The application of a
model in which the ionized gas covers only partially the X-ray source
did not improve the fit significantly ($\Delta\chi^2$ = 5 for one
additional parameter) and resulted in best-fit ionization parameter
$\xi$ consistent with the minimum allowed value of 10$^{-4}$ \ergsc.
We therefore consider the  cold absorber model as the
most appropriate description of the \chandra~spectrum of \ftm.

\begin{figure*}
\begin{center}
\includegraphics[width=6.5cm,height=8.5cm,angle=-90]{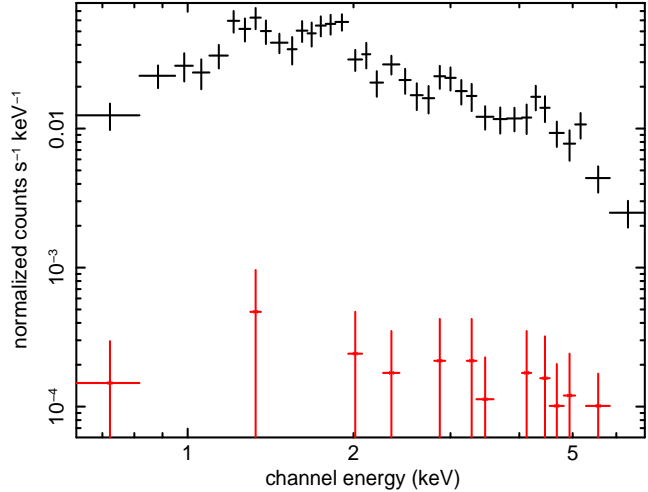}
\caption{The 9 ks \chandra~ACIS-S spectrum of \ftm~({\it top}) compared with the spectrum of the background ({\it bottom}). 
Over the \chandra~bandpass, the intensity of the source is more than an order of magnitude higher than background.}
\label{fig:56}
\end{center}
\end{figure*}

The data/model residuals for this model suggested a possible unmodelled
absorption feature at   1.52 keV (i.e. 2.15 keV in the quasar
rest-frame).  The addition of an absorption
Gaussian line in the fit resulted significant  at 93\% confidence
level.  We also checked the possibility that this line at
2.15$\pm$0.07 keV (EW=65$\pm$44 eV) could be an artifact owing to a bad
background subtraction. In particular, two well-known emission lines
from Al K$\alpha$ and Si  K$\alpha$  are present in the ACIS
background spectrum at 1.49 and 1.74 keV, respectively. However, the
background spectrum does not show any spectral feature and its
intensity is an order of magnitude fainter than the
background-subtracted signal from \ftm, e.g. Fig.~\ref{fig:56}. The
presence of the absorption line at $\sim$2.15 keV is very puzzling as
this energy does not  correspond to any   obvious line of appreciable
strength.  It can be associated to Si XIII-XV transitions  but, in
this case, much more prominent features from more abundant, lighter
elements are expected, which are instead not observed.  On the basis of
these arguments and the low statistical significance of this line, the
presence of this feature will be ignored in the following discussion.

Furthermore a small positive excess was seen around 4.5 keV
(i.e. $\sim$6.4 keV rest-frame).  The inclusion of a narrow Gaussian
emission line  in the NP model produced a statistical improvement of
$\Delta\chi^2$ = 3 for two additional parameters, corresponding to an
F-statistics probability of $\sim$90\%. We measured the centroid of
the line at 6.17$^{+0.11}_{-0.17}$ keV and an EW value of 232$\pm$175
eV.  Even if the energy of this line is not fully consistent with 6.4
keV,  an explanation in terms of \feka~emission is very likely. 

Finally, from this \chandra~observation we derived a 2-10(0.5-2) keV
flux of 1.3(0.2) $\times$ 10$^{-12}$ \cgs, corresponding to a
luminosity of 8(4.3)  $\times$ 10$^{44}$ \ergs.

\section{Broad-Band Spectral Properties of  \ftm}
\label{broadband}

A widely adopted way to evaluate the broad-band properties of AGN consists in 
estimating the optical UV-to-X-ray power-law slope \aox\footnote{
$\alpha_{\rm ox}=\frac{\log(f_{\rm 2~keV}/f_{2500~\mbox{\rm \scriptsize\AA}})}
{\log(\nu_{\rm 2~keV}/\nu_{2500~\mbox{\rm \scriptsize\AA}})}$, where 
$f_{\rm 2~keV}$ and $f_{2500~\mbox{\scriptsize \rm \AA}}$ are the 
rest-frame flux densities at 2~keV and 2500~\AA,
 respectively \citep[e.g.][]{avni80,zamorani81}} and comparing it with the values 
obtained from large samples of optically selected broad-line quasars 
\citep[e.g.,][]{vignali03b, strateva05,steffen06}. Unlike {\it blue} broad-line AGNs, which are usually 
characterized by low extinction, the optical/UV spectral properties of our 
target are consistent with heavy extinction \citep[U05;][]{urrutia08}. 

To obtain an ``intrinsic'' value for \aox, we applied a correction to the 
observed rest-frame 2500~\AA\ and 2~keV flux densities.
In the UV band, to derive the  luminosity at 2500~\AA, we used the magnitudes from the SDSS 
and adopted a procedure as described in \citet{vignali03b}. 
Then we assumed the nuclear extinction estimated by \citet{urrutia08} 
using HST data, i.e. E(B$-$V) = 1.29$\pm{0.12}$, coupled with an  SMC extinction curve with  $R_V=3.1$ as in \citet{urrutia08}
 to correct the   $L$(2500 \AA). 
In the X-rays, the correction for obscuration has been obtained 
directly from the best-fitting column density value (see Sect. \ref{fluxes}). 
The resulting \aox\ is $\approx-2.3$; at face value, this value is significantly steeper than that 
expected using the \aox\ vs. $L$(2500 \AA) correlation (e.g. Eq. (2) in \citet{steffen06}; see also \citet{just07}).
Such extreme values are more typical of the X-ray weak quasar population \citep{brandt00}, which is largely dominated by 
Broad Absorption Line (BAL) quasars \citep[e.g.][]{weymann91}.
Unfortunately, given the source redshift current optical spectral 
data are not able to  provide a direct observational evidence for
a possible BAL nature of \ftm~(e.g. Sect.~\ref{bal}).
 Steep \aox\ values can also explained in terms of  high Eddington ratios ($L_{\rm bol}/L_{Edd}$) as proposed by
\citet{kura09b} on
the basis of the predictions from the accretion disk/corona model of \citet{witt97}. According to
these authors high values of the accretion rate should lead to an
increase of the ratio between big blue bump and X-ray emission.
In this respect, observational evidence supporting the existence of a correlation between \aox\ and $L_{\rm bol}/L_{Edd}$  
comes from recent works based on large samples of AGNs \citep{young09,lusso09}.
This may bolster the hypothesis of \ftm\ being a BAL quasar since high $L_{\rm bol}/L_{Edd}$ values are typical of Narrow Line Seyfert 1 galaxies and BAL quasars, 
but  both optical (i.e. emission line
widths \simgt\ 1000 km s$^{-1}$) and X-ray spectral properties (i.e. relatively flat power-law photon index $\Gamma$
$\approx$ 1.65) of \ftm\ are not consistent with those of a  Narrow Line Seyfert 1 galaxy \citep[e.g.,][and
references therein]{leighly99}.

\section{Discussion}

The combination of X-ray brightness and a net exposure of $\sim$40 ks
makes the \xmm~spectrum of \ftm~one of the best exposed and most
detailed X-ray spectra of a \red~(i.e.  with $J-K_s>$1.7 and $R-K_s>$4.0) broad-line quasar to date.
This observation has definitely confirmed the complexity of the X-ray
spectrum of this luminous quasar.  In particular, our analysis has
revealed (i) the presence of a heavy  absorption with
\nh~$\approx$10$^{22}$ \cm2\ along the line of
sight to the nucleus  and (ii) the significant detection of a Compton
reflection component accompanied by a clear-cut \feka~emission line in
this AGN with \lum~$\approx$ 5 $\times$ 10$^{44}$ \ergs. 

The results from
re-analysis of the \chandra~data taken in 2004 are consistent with
those derived from the \xmm~observation. 
One immediate conclusion that can be drawn from these findings is that the
X-ray spectrum of \ftm~significantly differs from that typically
observed for optically-selected, broad-line quasars with comparable 2-10 keV
luminosity \citep{pico05,george00,just07}.  

\subsection{The nature of the X-ray absorber}

 The present X-ray observation is not able to provide unambiguous information about
 the location and the physical state of the absorber.
In fact, both a model with neutral and  mildly-ionized obscuring
material provide  an excellent fit to the \epic\ data (see Table
2). Unfortunately, no meaningful RGS data were collected for \ftm\ during this \xmm\ observation.

 In this Section, we discuss possible scenarios for the nature of the X-ray absorber in \ftm.
A direct statistical comparison between model NP and model WP favors
the warm absorption scenario at 99.6\% confidence level using the $F-$test.
Nonetheless,  when  a Compton reflection component with $R$=1 is added
to the NP model, the fit involving an ionized absorber is no longer
statistically preferable  to a model with cold absorption (i.e. model
refl-NP) as the difference in terms of $\Delta\chi^2$=1 for 1 dof is
significant at only 70\% confidence level according to the $F-$test. 

\subsubsection{Cold Absorption Scenario} 

\citet[][2005]{wilkes02} and U05 have indeed reported the common presence of 
cold absorption in the 
X-ray spectra of broad-line {\it red} quasars unlike
optically-selected objects \citep{pico05}. The value of
\nh$\sim$2 $\times$ 10$^{22}$ \cm2~measured for \ftm~(e.g. model
refl-NP in Table 2) is consistent with the distribution of the column
density  peaking around a few $\times$ 10$^{22}$ \cm2~observed for the
samples of \red\ 2MASS quasars.

Using a reddening of E(B-V)=1.29$\pm$0.12 mag ($A_V$$\approx$4 mag)
from \citet{urrutia08} and assuming for the
dust:gas ratio the average Galactic value of 1.7 mag cm$^{2}$/atoms
\citep[][]{bohlin78} we derived a \nh~value of 7.5$\pm$0.7 $\times$
10$^{21}$ \cm2, which is only a factor of $\approx$2.6 lower than the
\nh~measured from our \xmm~observation. This is interesting since
\citet{maiolino01} found that only low-luminosity
(i.e. \lum$<$10$^{42}$ \ergs) AGNs usually show a  E(B-V)/\nh~ratio
roughly consistent with the Galactic one, while \ftm~is a very
luminous quasar with  \lum~ $\approx$ 5 $\times$ 10$^{44}$ \ergs.  The
match between the column density measured via X-ray spectroscopy and
that expected on the basis of reddening E(B-V) due to the dust
suggests the likely possibility that the same material is responsible
for  the obscuration  both in optical and X-rays.  Similar findings
have been reported by \citet{kura09a} by an extensive study of the SED
of 44  \red, 2MASS-selected  AGNs at low $z$.  They found a substantial
agreement among the column densities inferred from the X-ray spectral
fitting and those derived from optical/IR colors once a very
detailed modeling of the optical colors is performed (i.e. including
the  combined effects due to  reddening and the contributions from the
AGN scattered unreddened light and the host galaxy  emission).  Using
Principal Component Analysis of the SED and emission-line parameters
for the same sample, \citet{kura09b} argue that the optical/X-ray
absorber may lie far from the nucleus, and presumably outside the
NLR. The interstellar medium   of the host galaxy with a
dust:gas ratio being comparable with Galactic dust might represent a
good candidate for this obscuration \citep{weingartner02, dai09}.  In
particular, \citet{kura09b}  proposed that obscuration possibly occurs
in the ISM of a host galaxy  inclined 
to our line of sight (LOS), although obscuration by circumnuclear dust is also important.

As described in Sect. \ref{complex}, our spectral analysis results
are also consistent with a scenario wherein the  absorber only
partially covers the X-ray emitting region. This  implies two distinct
LOS and that the obscuring screen along our  LOS is not uniform, being
spatially located in the nuclear region. The obscuring material must
indeed be close to the continuum source for partial covering to work.
A clumpy  (sub-)pc-scale {\it torus} as that proposed by
\citet{risaliti05} and \citet[][and references therein]{elitzur08} can
easily satisfy both these conditions. 
 The short time-scales
(i.e. few hours)
temporal variability of the absorbing column density observed in some Seyfert 2 galaxies
strongly supports this interpretation, for which
  the obscuring
matter is not continuously distributed and doughnut-shaped
\citep[e.g.][]{pier93,granato94} but consists of a number of clouds
distributed in order that the probability for direct viewing the AGN
increases away from the equatorial plane.  Broad-line emitting clouds
lying inside the dust sublimation radius ($R_d$) are dust-free and
extinguish only X-rays, while clouds being farther than  $R_d$ are
dusty and absorb the AGN light both at X-rays and optical
wavelength. 

However,  both the non-variability of the \nh~over the \xmm\ exposure time
 (also compared with the 2004 \chandra\ observation) 
and the consistency
between the \nh~values obtained from optical reddening and X-ray
analysis suggests that our LOS mostly intercepts dusty material
outside the $R_d$ (i.e. pc- and kpc-scale dust lanes and the interstellar medium in
the host galaxy), decreasing the probability it can partially cover
the X-ray nuclear source in \ftm.

\subsubsection{Ionized Absorption Scenario}
\label{ionized}
 
We found that a warm absorber model can also provide
an excellent description of the X-ray spectrum of \ftm.  Features from
ionized absorbers are detected in the UV and X-ray spectra of over
half of AGNs \citep{crenshaw99,reynolds97, pico05,mckernan07} which
are interpreted as the signature of an outflowing multi-component wind
rising vertically from the accretion disk intercepting our LOS
\citep[][and references therein]{crenshaw03, krongold07}.  The large
deficit below $\sim$2 keV visible in  Fig. 1 (left panel) would be the result 
of the interplay between large column
density and moderate ionization level of the obscuring gas. Such a
combination of physical parameters of the warm absorber is not
commonly observed in unreddened Type 1 AGNs \citep[e.g.][]{mckernan07}
and it could be responsible for the classification of \ftm~as
\red~quasar.  However, our \epic\ data cannot provide the adequate
resolution and sensitivity to significantly detect the likely presence
of multiple ionization components and, therefore, the measured values
of \nh~and $\xi$  must be considered as average values of these
parameters.  Recent observations  \citep{steen09,krongold07} have
pinpointed the distance of absorption components of the warm absorber
in AGNs at sub-pc scale from the X-ray source. The clumpy and
outflowing nature of the AGN winds  lends support to a
partial-covering scenario, which is consistent with our spectral analysis results (e.g. Sect. \ref{complex}), 
with two different LOS, one of which does not pass through the wind. 

As described in Sect.~\ref{complex} we also tested a model wherein
the soft X-ray extra-continuum
component is explained in terms of  a mix of scattered emission and photoionized emission from the
 moderately-ionized absorber itself (e.g. model refl-wabs in Table 2). 
We yielded a very good fit to the
\xmm~spectral data with $\xi$$\approx$6 \ergsc~and a column density of
\nh~$\approx$ 1.3 $\times$ 10$^{22}$ \cm2~for the obscuring gas. These
values indicate a  low ionization state of the absorber that
significantly reduces the  primary AGN emission in the soft X-ray
portion of the spectrum and, in turn, could unveil weak emission
components which are usually overwhelmed by the continuum in normal
broad-line AGNs.  Interestingly, a similar scenario was also suggested to
explain the {\it soft-excess} emission in the \red\ quasars 2MASS
23449$+$1221 \citep{pounds05}, 2MASS 1049$+$5837  \citep{wilkes08} and 2MASS 0918$+$2117 \citep{pounds07},
and, furthermore,   for the Seyfert 1 galaxy Mkn 335 during a very low
continuum state \citep{longinotti08}.\\

Finally, if a cold absorption component is added to the warm absorber
fitting models, no improvement in the fit statistic is yielded,
with an upper limit of \nh $<$6 $\times$ 10$^{20}$ \cm2 on its column
density, indicating that a scenario with an innermost highly ionized
absorber embedded in large-scale (i.e. galactic) neutral absorber
cannot explain the X-ray spectrum of \ftm.

\subsection{FTM 0830$+$3759 as a possible BAL  quasar candidate}
\label{bal}

The presence of a nuclear outflow in \ftm~can be interesting in the
light of the results recently obtained by \citet{urrutia09} for a
spectroscopic survey of \red~quasars with selection criteria very
similar to those used in  the FTM survey.  For this sample they found
a significantly larger fraction (\simgt40\%) of 
BAL quasars at $z$ $>$ 0.9 than that observed in  optical surveys \citep{hewett03}.
BALs are interpreted as direct evidence of a mass outflow from the AGN
\citep{weymann91,elvis00,hasinger02,young07}.  The frequent detection
of BALs in \red~quasars  leads   \citet{urrutia09} to suggest that BAL
quasars may represent a peculiar dust-enshrouded  phase in the evolution of  quasars.
  The appearance as a BAL quasar might indeed be linked to  the duty cycle of an accreting supermassive black
hole instead of being simply due to an orientation effect and the
geometry of the outflow.  Moreover their powerful nuclear winds may
be the main mechanism responsible for the AGN-driven   feedback
thought to be relevant for star formation quenching and invoked in
many models of joint formation and co-evolution of quasars and their massive
spheroidal host galaxies  \citep{hopkins06,dimatteo05,silk98}.  Evidence for a large
fraction of mergers and ongoing galaxy interactions, that are believed
to be the trigger of both starburst and AGN activity  at the very
initial stages of galaxy/AGN co-evolution has been indeed observed
amongst host galaxies of FTM luminous quasars by \citet{urrutia08}.
\citet{georgakakis09}  have recently discovered that the level of
star-formation activity in the luminous, high-$z$, \red~quasars is higher than in 
optically-selected objects. This provides further clues to an
evolutionary scenario where \red~quasars are young, dust-embedded
objects in which the AGN driven feedback processes have not yet
completely inhibited the star-formation in their host galaxies. 
As described in Sect.~\ref{broadband}, the steep \aox\ value (\aox\ = $-$2.3) reported for \ftm\
 is favorably compatible
 with the observed broad-band properties of BAL quasars
and, furthermore, the complex, absorption-dominated  spectrum of \ftm~resembles the typical
 X-ray spectrum of a BAL quasar \citep[e.g.][]{gallagher02,schartel05}.
Unfortunately the redshift of \ftm~is not in the range suited to provide a direct BAL classification for this
 quasar by a ESI/Keck spectrum as pointed out by \citet{urrutia09}. Future observations will be able to
 detect the possible absorption troughs in the optical-UV spectrum of this reddened quasar in order to 
shed light on the presence of a strong outflowing nuclear wind.

\subsection{Comparison with the Hard X-ray Properties of unreddened Broad-line Quasars}

Our analysis of the \xmm~observation of \ftm~has revealed the presence
of the typical features of a Comptonized reflected spectrum from a
neutral material, i.e.  an intense fluorescent \feka~emission line and
a hard X-ray Compton reflection component
\citep{george91,krolik94,perola02}.  Given the paucity of well-exposed
quasars with \lum\simgt 5 $\times$ 10$^{44}$ \ergs~information
regarding the presence and the properties of reflection features in
their hard X-ray spectra are still sparse \citep[see
also][]{miniutti09}. In particular, since the peak of the Compton hump
occurs at $\sim$30 keV,  the sensitivity reached by current and past
X-ray telescopes observing in this energy band allows to efficiently
study only very few bright sources, which are predominantly local
Seyfert galaxies. 

The line equivalent width measured with respect to the reflected
continuum assuming $R$=1 is EW $\sim$ 800 eV, that is consistent with
the theoretical prediction of EW $\sim$ 1000 eV for reflection from
optically thick material with iron solar abundance
\citep[e.g.][]{matt96} suggesting an origin in the reflecting medium.
Bearing in mind the uncertainty in the exact value of $R$, this in
turn implies the presence of off-LOS Compton-thick material in this
quasar that can be likely associated with  a pc-scale structure
\citep{jaffe04,meisen07} as the putative {\it torus} envisaged by the
AGN Unified Model \citep{antonucci93}.  On the other hand, the value
of the EW calculated with respect to the unabsorbed primary continuum
is $\sim$160  eV, that is significantly  higher than the
predicted EW  from a uniform shell of material with  a
column density of \nh~$\sim$ 2 $\times$ 10$^{22}$ \cm2~encompassing
the continuum source, i.e. EW $<$ 40 eV \citep{guainazzi05, matt02}.
Unreddened broad-lined quasars at the same hard X-ray luminosity level
of \ftm~typically show an \feka~line with an $\langle$EW$\rangle$ in
the range from 40 eV \citep{bianchi07b} to 70 eV
\citep{jimenez05}.  The large EW value reported for \ftm\ can be explained
with a large covering fraction of the Fe-emitting material in this \red\ quasar or,
alternatively, it could be
associated to a low-flux state of the  central source, with the 
distant reflector still responding to an  average higher luminosity.

Finally, once a reflection continuum is included in the fitting model,
the slope of the power-law continuum measured for \ftm~($\Gamma$
$\approx$1.65) is still slightly flatter than the canonical spectral
index measured for classical radio-quiet  quasars, i.e 1.8--1.9
\citep{just07,pico05}.  A low-flux state can also be invoked for
interpreting the observed slope \citep{grupe08}.  Furthermore, the value of $\Gamma$
$\approx$1.65 is at odds with the claim of U05 that \red~quasars show
an average  steep photon index of $\langle\Gamma\rangle$=2.1--2.2.
However such conclusion is based on values of $\Gamma$ inferred from
low-count spectra or hardness-ratio analysis and therefore affected by
large uncertainties: deeper X-ray follow-up observations of these
\red~quasars indeed found spectral index values consistent with those
typical of optically-selected quasars
\citep[e.g.,][]{wilkes05,pounds05}.\\

\acknowledgements
We thank the anonymous referee for valuable feedback.
EP is grateful to R.~Maiolino, M.  Giustini, L. Ballo and M. Mignoli for helpful discussions and suggestions.
EP, CV and SB acknowledge support under ASI/INAF contract I/088/06/0.
FN acknowledges the XMM-Newton grant NNX08AY05G. GM thanks the Ministerio de Ciencia e Innovaci\'on and CSIC for support 
through a Ram\'on y Cajal
contract.
 This research has made use of the SIMBAD database, operated at CDS (Strasbourg, France) and the NASA/IPAC Extragalactic Database 
 (NED) which is operated by the Jet Propulsion Laboratory, California Institute of Technology, under contract with the National Aeronautics and Space Administration. 
 Based on observations obtained with XMM-Newton, an ESA science mission with instruments and contributions directly funded by ESA Member States and NASA. 
%

{}
\end{document}